\documentclass[a4paper,12pt]{article}
\usepackage{graphicx,rotating,hyperref,slashed,amsmath,charter,xcolor,catchfilebetweentags,ifluatex,cite}
\usepackage{soul}
\makeatletter
\hypersetup{colorlinks,bookmarksopen,bookmarksnumbered,
linkcolor=blus,pdfstartview=FitH,urlcolor=rossos,citecolor=verde}

 \newcommand{\bp}{\bar M_{\rm Pl}}
 \renewcommand{\Im}{{\rm Im}\,}

\newcommand{\mio}[1]{}

\newcommand{\fig}[1]{~\ref{fig:#1}}
\newcommand{\gs}{f_0}
\newcommand{\gt}{f_2}

\definecolor{Gray}{gray}{0.95}

\newcommand{\sfrac}[2]{#1/#2}

\usepackage{multicol}
\usepackage{color}
\usepackage{amssymb}
\definecolor{rosso}{cmyk}{0,1,1,0.4}
\definecolor{rossos}{cmyk}{0,1,1,0.55}
\definecolor{rossoc}{cmyk}{0,1,1,0.2}
\definecolor{blu}{cmyk}{1,1,0,0.3}
\definecolor{blus}{cmyk}{1,1,0,0.6}
\definecolor{bluc}{cmyk}{1,1,0,0.1}
\definecolor{verde}{cmyk}{0.92,0,0.59,0.25}
\definecolor{verdec}{cmyk}{0.92,0,0.59,0.15}
\definecolor{verdes}{cmyk}{0.92,0,0.59,0.4}

\newcommand{\eq}[1]{~{\rm (\ref{eq:#1})}}

\newcommand{\GeV}{\,{\rm GeV}}

\def\circa#1{\,\raise.3ex\hbox{$#1$\kern-.75em\lower1ex\hbox{$\sim$}}\,}

\newcommand{\beq}{\begin{equation}}
\newcommand{\eeq}{\end{equation}}

\newcommand{\bea}{\begin{eqnarray}}
\newcommand{\eea}{\end{eqnarray}}
\newcommand{\be}{\begin{equation}}
\newcommand{\ee}{\end{equation}}
\font\tenrsfs=rsfs10 at 12pt
\font\sevenrsfs=rsfs7 at 10 pt
\font\fiversfs=rsfs5
\newfam\rsfsfam
\textfont\rsfsfam=\tenrsfs
\scriptfont\rsfsfam=\sevenrsfs
\scriptscriptfont\rsfsfam=\fiversfs
\def\mathscr#1{{\fam\rsfsfam\relax#1}}

\def\Lag{\mathscr{L}}

\def\circa#1{\,\raise.3ex\hbox{$#1$\kern-.75em\lower1ex\hbox{$\sim$}}\,}
\makeatletter

\def\hhref#1{\href{http://arxiv.org/abs/#1}{arXiv:#1}} 

\font\ital=cmu10 

\def\hhref#1{\href{http://arxiv.org/abs/#1}{arXiv:#1}} 
 
\def\art{\@ifnextchar[{\eart}{\oart}}
\def\eart[#1]#2#3#4#5#6{{\rm #2}, {\em #3 \bf #4} {\rm (#6) #5} ({\em #1})}

\def\article{\@ifnextchar[{\earticle}{\oarticle}}
\def\oarticle#1#2#3#4#5#6{{\rm #1}, {\ital ``#6''}, {\rm #2 #3 (#5) #4}}
\def\earticle[#1]#2#3#4#5#6#7{{\rm #2}, {\ital ``#7''}, {\rm #3 #4 (#6) #5}  [\hhref{#1}]}
\def\hepart[#1]#2{{\rm #2, \ital#1}}
\def\heparticle[#1]#2#3{#2, {\ital ``#3''} [\hhref{#1}]}

\newcounter{alphaequation}[equation]
\def\thealphaequation{\theequation\hbox to
0.6em{\hfil\alph{alphaequation}\hfil}}
\def\eqnsystem#1{
\def\@eqnnum{{\rm (\thealphaequation)}}
\def\@@eqncr{\let\@tempa\relax \ifcase\@eqcnt \def\@tempa{& & &} \or
  \def\@tempa{& &}\or \def\@tempa{&}\fi\@tempa
  \if@eqnsw\@eqnnum\refstepcounter{alphaequation}\fi
\global\@eqnswtrue\global\@eqcnt=0\cr}
\refstepcounter{equation} \let\@currentlabel\theequation \def\@tempb{#1}
\ifx\@tempb\empty\else\label{#1}\fi
\refstepcounter{alphaequation}
\let\@currentlabel\thealphaequation
\global\@eqnswtrue\global\@eqcnt=0 \tabskip\@centering\let\\=\@eqncr
$$\halign to \displaywidth\bgroup \@eqnsel\hskip\@centering
$\displaystyle\tabskip\z@{##}$&\global\@eqcnt\@ne
\hskip2\arraycolsep\hfil${##}$\hfil& \global\@eqcnt\tw@\hskip2\arraycolsep
$\displaystyle\tabskip\z@{##}$\hfil
\tabskip\@centering&\llap{##}\tabskip\z@\cr}
\def\endeqnsystem{\@@eqncr\egroup$$\global\@ignoretrue} \makeatother

\newcommand{\eV}{\,{\rm eV}}

\begin{document}
\centerline{CERN-TH-2018-195}

\vspace{0.2cm}

\begin{center}
{\Huge \bf \color{rossos} New infra-red enhancements\\[3mm] in 4-derivative gravity}\\[1cm]

{\large\bf Alberto Salvio$^{a}$, Alessandro Strumia$^{a,b}$  {\rm and} Hardi Veerm\"{a}e$^{a,c}$}  
\\[7mm]
{\it $^a$ } {\em CERN, Theoretical Physics Department, Geneva, Switzerland}\\[3mm]
{\it $^b$} {\em Dipartimento di Fisica dell'Universit{\`a} di Pisa and INFN, Italia}\\[3mm]
{\it $^c$} {\em National Institute of Chemical Physics and Biophysics, Tallinn, Estonia}

\vspace{1cm}
{\large\bf\color{blus} Abstract}
\begin{quote}\large
4-derivative gravity  provides a renormalizable theory of quantum gravity at the price
of introducing a physical ghost, which could admit a sensible positive-energy quantization. 
To understand its physics, we compute ghost-mediated scatterings among matter particles at tree-level,
finding a  new power-like infra-red enhancement typical of 4-derivative theories, that we dub
`ghostrahlung'.
Super-Planck\-ian scatterings get downgraded to Planckian
by radiating hard gravitons and ghosts, 
which are weakly coupled and
carry away the energy.


\end{quote}

\thispagestyle{empty}
\end{center}
\begin{quote}
{\large\noindent\color{blus} 
}

\end{quote}
\tableofcontents

\setcounter{footnote}{0}



\section{Introduction}
\begin{figure}
\begin{center}
\includegraphics[width=0.95\textwidth]{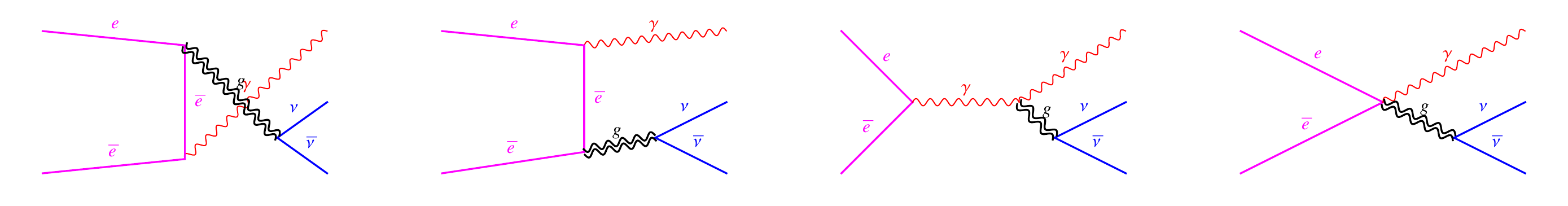}
\caption{\em Sample of gravi-ghost-mediated $2\to 3$ process:
the differential distribution of
$\sigma(e\bar e\to \gamma\nu\bar\nu)$ is used to infer the 
cross section for $\sigma(e\bar e\to \gamma g)$.
\label{fig:Feyn23}}
\end{center}
\end{figure}

4-derivative gravity provides a renormalizable gravity theory~\cite{Stelle:1976gc},
controlled by dimensionless coupling constants, called $f_2$ and $f_0$ in~\cite{agravity}.
In the presence of the dimension-full super-renormalizable Einstein term, 
the graviton splits into $g$ (the massless graviton) and
$g_2$ (a spin-2 state with mass $M_2$; an extra spin 0 component $g_0$ with mass $M_0$ is less relevant)
as clear from the decomposition of its propagator, with Lorentz indeces omitted
\beq\label{eq:p42p2}P(k^2)=
\frac{1}{M^2_2 k^2 - k^4} = \frac{1}{M^2_2} \bigg[\frac{1}{k^2} - \frac{1}{k^2 - M^2_2}\bigg].\eeq
This cancellation between $g$ and $g_2$ in the virtual propagator makes gravity renormalizable;
however, the minus sign means that $g_2$ classically has negative kinetic energy.
The same problem was encountered with classical fermions.
4-derivative theories too admit a positive-energy quantization, but at the price
of an indefinite quantum norm  that obscures the probabilistic interpretation~\cite{Salvio:2015gsi,hep-th/0608154,Raidal:2016wop,Strumia:2017dvt,Salvio:2018crh}. 
In view of this situation, we here explore how the ghost behaves 
making two pragmatic simplifications.
\begin{itemize}
\item[1.]
First, we restrict the attention to observables
measurable from asymptotic distance: life-times and cross sections.
A ghost is then indirectly defined through its effects as an intermediate virtual particle in 
Feynman diagrams that describe scatterings among matter particles
(scalars, fermions and vectors). 

\end{itemize}
This is how collider experiments
reconstruct any short-lived particle from kinematical distributions of
final-state particles.
When the intermediate particle is a ghost, this  is known as `Lee-Wick approach'~\cite{LW,Coleman},
and  ambiguities appear at higher orders,
in diagrams that probe configurations with two ghosts.
\begin{itemize}
\item[2.]
Second, we focus on tree-level processes, 
not affected by higher order ambiguities,
and that can probe the generic gravi-ghost kinematics.
Fig.~\ref{fig:Feyn23} shows an example of this.
\end{itemize}
Under these assumptions, we will extract a good deal of ghost physics, 
common to various attempts of fully defining the theory.
The theory is well defined in the Euclidean space:
according to~\cite{1703.04584} a generalization of the Wick rotation 
defines Minkowskian physics solving the above-mentioned ambiguities perturbatively to all orders.
When evaluating tree-level diagrams their approach reduces to ours:
integrated cross sections are not affected by the extra structure
assumed in~\cite{1703.04584} on the top of the ghost resonance;
all momenta that enter our expressions are Minkowskian physical observables described by real numbers.
The other approaches that extract probabilities from negative norms~\cite{hep-th/0608154,Strumia:2017dvt,Salvio:2018crh}
similarly reduce to our results when evaluated at leading order in the couplings.


\smallskip

Cross sections in QED and QCD are affected by soft and collinear infra-red (IR) divergences.
These effects are well understood thanks to soft theorems~\cite{vonWeizsacker:1934nji,Williams:1934ad,Altarelli:1977zs},  which also apply to Einstein gravity~\cite{Weinberg:1965nx,1109.0270,1207.4926}.
We find a new kind of IR enhancement related to the 4-derivative structure
and to the consequent gravi-ghost propagator of Eq.\eq{p42p2}.
Cross sections mediated by a gravi-ghost contain a factor $f_2^2 s  \int d(k^2) \, P(k^2)$.
In the limit of massless gravi-ghost this factor is power IR divergent at small $k^2$.
As a result,  cross sections such as $e^+e^-\to \gamma \nu\bar\nu$ (fig.~\ref{fig:Feyn23}) 
do not have the form expected in theories with dimensionless coupling,
$\sigma \sim 1/s$ times powers of the couplings.
Indeed,  in the massless limit, the Newton potential $V \propto 1/r$
gets replaced by a confining $V \propto r$: free particles disappear in this limit.

In the realistic massive theory, the IR divergence is cut by the gravi-ghost mass $M_2$.
Then each massive gravi-ghost contributes to the cross section with a multiplicative enhancement
$ f_2^2 s/M_2^2 \sim s/M_{\rm Pl}^2$.
Gravi-ghost radiation becomes an order one correction in super-Planckian collisions.
While purely gravitational cross sections can remain smaller than in Einstein theory,
cross sections mediated by large matter couplings 
(such as fig.~\ref{fig:Feyn23})
look as bad as those in UV-divergent Einstein theory, which violate naive perturbative unitarity.
This reassures that ghosts do not do miracles, like
cancelling positive cross sections with negative cross sections.

\smallskip

In agravity cross sections grow because of the new IR enhancement of gravi-ghost emission.
Unlike in Einstein gravity, Planckian gravitons negligibly interact and simply carry away their energy.
As a result energies above the Planck scale get radiated down to sub-Planckian energies
without forming, at the same time, non-perturbative structures such as black holes.



%
%
%
%


%


\smallskip



\medskip

%
%
%

An additional issue is that the IR enhancement is saturated at the ghost pole.
To understand what it is, one needs to go beyond perturbation theory.
As well known, perturbative corrections diverge when an intermediate particle goes on-shell.
A non-perturbative resummation transforms a matter pole into a Breit-Wigner peak,
$1/(k^2 - m^2 + i k \Gamma)$.

A matter particle acquires a decay width $\Gamma$ with the same sign
as the $i \epsilon$  prescription in its Feynman propagator, $1/(k^2-m^2 + i \epsilon)$,
which defines the theory as the continuation from the Euclidean known as `Wick rotation'.
  On the other hand,
a massive ghost that decays into matter particles
acquires a negative decay width $\Gamma<0$, 
which makes its behaviour acausal on microscopic scales~\cite{Coleman}.

\medskip

The paper is structured as follows.
In section~\ref{rates} we derive rates of generic gravi-ghost-mediated processes.
In section~\ref{specific} we compute specific examples.
In section~\ref{Ghostrahlung} we interpret IR-enhanced rates.
Conclusions are given in section~\ref{end}.

\section{Rates for gravi-ghost-mediated processes}\label{rates}

\subsection{Summary of the theory and of notations}
Following~\cite{Stelle:1976gc} we consider the renormalizable action (in the notation of~\cite{agravity})
\beq \label{eq:Ladim}
S=\int d^4x\,\sqrt{|\det g|} \bigg[ \frac{R^2}{6\gs^2} + \frac{\frac13 R^2 -  R_{\mu\nu}^2}{\gt^2} -\frac{\bp^2}{2}R
 + \Lag_{\rm matter}\bigg],
\eeq
where  the first two terms, suppressed by the dimensionless gravitational couplings $f_0$ and $f_2$,
are the 4-derivative graviton kinetic terms;
the latter term $\Lag_{\rm matter}$ is the part of the Lagrangian that depends on the matter fields
(scalars $S$, fermions $\psi$, vectors $V$ with gauge-covariant kinetic terms, Yukawa couplings, quartic scalars, 
scalar couplings to gravity, $-\xi_S |S|^2 R$ and possibly with super-renormalizable terms).

The Einstein-Hilbert term, in the middle, could be induced dynamically from a dimensionless action, e.g.~as $\bp^2/2=\xi_S\langle S\rangle^2$~\cite{agravity}.
In its presence, the 4-derivative graviton splits into the massless spin-2 graviton, a spin-2 ghost $g_2$ with mass
$M_2 = f_2 \bp/\sqrt{2}$, a spin-0 scalar $g_0$ with mass $M_0 = f_0 \bp/\sqrt{2}$.
We collectively denote them as `gravi-ghost'. The gravi-ghost propagator is
(in the gauge $\xi_h=c_g=0$ of~\cite{agravity})
 \beq 
\label{eq:Pgrav}
 P_{\mu\nu\,\alpha\beta}(k^2) = i\sum_{j=\{0,2\}} c_j f_j^2 P^{(j)} (k^2) \Pi^{(j)}_{\mu\nu\alpha\beta}\, ,\eeq
where $k_\mu$ is the quadri-momentum, $c_0=1$, $c_2=-2$ (this sign is crucial)
and 
\beq \label{eq:Presum}
P^{(2)}(k^2) = \frac{1}{k^2(k^2-M_2^2 + i k \Gamma_2)},\qquad
P^{(0)}(k^2) = \frac{1}{k^2(k^2-M_0^2+ i k \Gamma_0)}.
\eeq
We added  decay widths $\Gamma_j$ which  appear after (naively) 
resumming quantum corrections to the propagator.
The $\Pi^{(j)}_{\mu\nu\alpha\beta}$ are orthogonal projectors over the spin components~\cite{Stelle:1976gc,agravity}, which sum up to unity:
 $(\Pi^{(2)}+\Pi^{(1)}+\Pi^{(0)}+\Pi^{(0w)})_{\mu\nu\alpha\beta}=\frac12 (\eta_{\mu\nu}\eta_{\alpha\beta}+\eta_{\mu\beta}\eta_{\alpha\nu})$. 
 We only need $\Pi^{(2)}$ and $\Pi^{(0)}$, which are 
\bea
\Pi^{(2)}_{\mu\nu\rho\sigma} &=& \frac12 T_{\mu\rho}T_{\nu\sigma} + \frac12 T_{\mu\sigma} T_{\nu\rho}-\frac{T_{\mu\nu}T_{\rho \sigma}}{3}, \label{P2def} \\
\Pi^{(0)}_{\mu\nu\rho\sigma} &=& \frac{T_{\mu\nu}T_{\rho \sigma}}{3}, \label{P0def} 
 \eea
where $T_{\mu\nu} = \eta_{\mu\nu} - k_\mu k_\nu/k^2$ is the transverse projector.

\begin{figure}
\begin{center}
\includegraphics[width=0.5\textwidth]{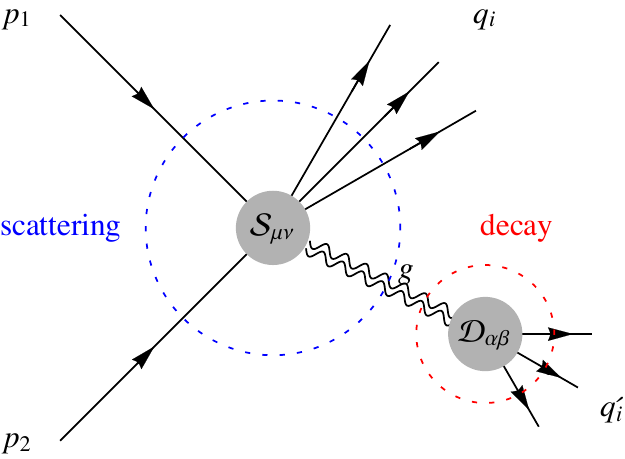}
\caption{\em \label{fig:FeynSD} A generic process among matter particles mediated by one gravi-ghost, denoted as a double wave.}
\end{center}
\end{figure}

\subsection{Processes mediated by one gravi-ghost}\label{1g}
In order to study the gravi-ghost behaviour
we consider a generic scattering between matter particles that contains one
intermediate virtual
gravi-ghost $g$ with quadri-momentum $k_\mu$ that transforms into matter particles with quadri-momenta $q'_i$, as illustrated in fig.\fig{FeynSD}.
We denote the quadri-momenta of the other final-state particles as $q_i$, 
and those of the two initial-state particles as $p_1$, $p_2$, such that momentum conservations reads 
$\sum p_i = \sum q_i + \sum q'_i$. 
We decompose the process as a `scattering' with momenta 
$\sum p_i = k+\sum q_i$ times a gravi-ghost `decay' with momenta $k = \sum q'_i$.
Concrete examples are $2\to 2$ scatterings such as $e(p_1) \bar e(p_2) \to g (k)\to \nu(q'_1)\bar\nu(q'_2)$
(where $k^2$ is fully determined by kinematics) and $2\to 3$ scatterings such as
$e(p_1)\bar e(p_2)\to \gamma(q_1)\nu(q'_1)\bar\nu(q'_2)$ as shown in fig.~\ref{fig:Feyn23}.

\medskip

The cross section is $d\sigma/d\Phi = |\mathscr{A}|^2/4I$ 
where $I\equiv \sqrt{(p_1\cdot p_2)^2 - m_1^2 m_2^2} $ is the usual flux factor
($I =s/2$ for massless particles with $s=(p_1+p_2)^2$) and 
$d\Phi$ is the usual relativistic phase space.
The cross section is  well defined despite the virtual gravi-ghost, 
which gets indirectly defined by what it does.
In order to proceed in understanding the ghost, we decompose
the scattering amplitude $\mathscr{A}$ 
as the amplitude $\mathscr{S}_{\mu\nu}$ for the `scattering' 
times the gravi-ghost propagator $ P_{\mu\nu\alpha\beta}$ of eq.\eq{Pgrav},
times the amplitude $\mathscr{D}_{\alpha\beta}$ for the `decay':
\beq \mathscr{A}= \mathscr{S}_{\mu\nu} P_{\mu\nu\alpha\beta}(k^2)  
\mathscr{D}_{\alpha\beta}.
\eeq
We show only the gravi-ghost indices, leaving implicit
the matter indices.
We also leave implicit the usual sum (average) over their initial-state (final-state) components:
polarizations, other multiplicities, etc.
Inserting $1 = \int d^4k~\delta(k- \sum q'_i)$ times
$1 = \int ds_{g} \,\delta (k^2-s_{g})$
the phase space decomposes as
\beq d\Phi = d\Phi_{\rm scattering} \frac{ds_{g}}{2\pi} d\Phi_{\rm decay}\, ,\eeq
where $d\Phi_{\rm decay}$ is the phase space for the decay of a particle with squared mass $s_g = k^2$
into particles with quadri-momenta $q'_i$, and 
$d\Phi_{\rm scattering}$ is the phase space for producing the gravi-ghost and the  final-state particles with momenta $q_i$.
If the set of final-state particles with momenta $q_i$ is empty,
$d\Phi_{\rm scattering}$ contains a Dirac $\delta$ which removes the $ds_g$ integral.

\medskip

In general, the different spin components of  intermediate particles give different angular distributions
of final-state particles,
but have the same decay width, because of Lorentz invariance.
Thereby the squared amplitude simplifies
 after integrating over $d\Phi_{\rm decay}$.
Since  the integrated squared decay amplitude  only depends on $k$, 
 and since $\mathscr{D}_{\alpha\beta} k_\alpha=0$, it has the form
\beq\label{eq:DDPi}
\int d\Phi_{\rm decay}   \mathscr{D}_{\alpha\beta}\mathscr{D}^*_{\alpha'\beta'}= 
\int d\Phi_{\rm decay} \sum_{j=\{0,2\}}|D^{(j)}|^2 \Pi^{(j)}_{\alpha\beta\alpha'\beta'} \, , \eeq
where
\beq
|D^{(2)}|^2 =\frac{1}{5} \Pi^{(2)}_{\alpha\beta\alpha'\beta'} \mathscr{D}_{\alpha\beta}\mathscr{D}^*_{\alpha'\beta'},\qquad
|D^{(0)}|^2 =
\Pi^{(0)}_{\alpha\beta\alpha'\beta'} \mathscr{D}_{\alpha\beta}\mathscr{D}^*_{\alpha'\beta'}.
\eeq
Since $\Pi^{(j)}$ are orthogonal projectors the total squared amplitude simplifies to
\beq
\int d\Phi\,   |\mathscr{A}|^2 = \int d\Phi \sum_{j=\{0,2\}}  |S^{(j)} D^{(j)} P^{(j)}|^2\, ,\eeq
where the squared `scattering' amplitude summed over gravi-ghost polarizations is 
\beq  |S^{(j)}|^2 \equiv  \mathscr{S}_{\mu\nu} \Pi^{(j)}_{\mu\nu\mu'\nu'} \mathscr{S}^*_{\mu'\nu'} .\eeq
The cross section splits into its `scattering' and `decay' parts as
\beq \label{eq:sigma23}{
\frac{d\sigma}{ d\Phi_{\rm scattering} ds_g} =
\sum_{j=\{0,2\}}   \frac{c_j f_j^2}{4\pi I }  
  |S^{(j)}|^2  \times | P^{(j)}(s_g)|^2
\times\frac{c_jf_j^2}{2}
\int d\Phi_{\rm decay} |D^{(j)}|^2}.\eeq
At $s_g\gg M^2_j$ the modulus squared of the 4-derivative propagator reduces to $1/s_g^4$: the graviton-graviton and the ghost-ghost terms get cancelled
by the ghost-graviton interference term, which vanishes on-shell at $s_g=M_2^2$.

\subsection{Cross section for producing $N$ gravi-ghosts}
The cross section for producing two on-shell gravi-ghosts
$g_{\mu_i\nu_i}$
with momenta $k_i$, squared masses
$s_i=k_i^2$, spin $j_i$, where $i=\{1,2\}$
can be analogously extracted from the on-shell part of 
scattering amplitudes among matter particles mediated by two gravi-ghosts.
The phase space  decomposes as
\beq d\Phi=\frac{ds_1}{2\pi}\frac{ds_2}{2\pi}
d\Phi_{\rm scattering} d\Phi_{\rm decay1}d\Phi_{\rm decay2}.\eeq
Proceeding analogously to the previous section, we define
as  $\mathscr{S}_{\mu_1\nu_1 \mu_2\nu_2}$ the `scattering' sub-amplitude,
such that the cross section for production of two on-shell gravi-ghosts is\footnote{Note the absence of a symmtery factor for identical intermediate gravitons $\sigma_{j_1 j_2}$. The appropriate symmetry factor should of course be included in the total cross section $\sigma(\ldots \to (g \to \ldots)(g \to \ldots ))$ in case the gravi-ghosts decay into identical particles.}
\beq
\sigma_{j_1 j_2} =
\frac{1}{4I}\frac{c_{j_1} f_{j_1}^2}{ M_{j_1}^2} 
\frac{c_{j_2} f_{j_2}^2}{ M_{j_2}^2}
 \int d\Phi_{\rm scattering}  |S^{(j_1,j_2)}|^2\, ,\eeq
where 
\beq
 |S^{(j_1,j_2)}|^2=
 \mathscr{S}_{\mu_1\nu_1 \mu_2\nu_2} 
\Pi^{(j_1)}_{\mu_1\nu_1\mu'_1\nu'_1}\Pi^{(j_2)}_{\mu_2\nu_2\mu'_2\nu'_2} \mathscr{S}^*_{\mu'_1\nu'_1\mu'_2\nu'_2}.
 \eeq
 Analogous expressions hold for $N$ gravi-ghosts.

\subsection{Processes mediated by one on-shell gravi-ghost}
As usual, the cross section is dominated by the phase-space region where
the gravi-ghost is on-shell, if this is kinematically allowed.
In the narrow-width approximation, around the poles at $s_g \simeq M_j^2$ the gravi-ghost squared propagator approximates as
\beq
|P^{(j)}|^2 \simeq \frac{\pi}{M_j^5|\Gamma_j|} \delta(k^2-M^2_j).
\eeq
Thereby Eq.\eq{sigma23} reduces on-shell to
\beq \label{eq:2->3}
\frac{d\sigma}{d\Phi_{\rm scattering}} \simeq \sum_{j=\{0,2\}}
\frac{d\sigma_{j} }{d\Phi_{\rm scattering}} \frac{\Gamma_{j\to f}}{ |\Gamma_j|},\eeq
where
\beq
\Gamma_{j\to f}=\frac{c_jf_j^2}{2M_j^3}
\int d\Phi_{\rm decay} |D^{(j)}|^2\label{eq:Gammaj}
\eeq
is the partial decay width of the component of the gravi-ghost with spin $j$
with total decay width $\Gamma_j$, and
\beq\label{eq:sigmagj}
{d\sigma_{g_j}= \frac{1}{4I}\frac{c_j f_j^2}{ M_j^2}   |S^{(j)}|^2}=
\frac{1}{4I}\frac{2c_j }{ \bp^2}  |S^{(j)}|^2
\eeq
is the differential cross section for gravi-ghost single production.
The factors $(c_j f_j/M_j^2)^2$ have been split symmetrically among scattering and decay, getting the standard normalization of decay widths in section~\ref{2->2}.
Although  $\Gamma_j$ and $d\sigma_j$ are negative when the ghost with $c_2<0$ is involved,
the cross section among matter particles $d\sigma$ is always positive.
The negative width signals microscopic acausality~\cite{Coleman}.

\medskip

The cross section for producing the massless graviton $g$ is 
\beq
d\sigma_{g}= \frac{1}{4I} \frac{2}{\bp^2}   |S^{(g)}|^2d\Phi_{\rm scattering}
\stackrel{s\gg M_{0,2}^2}\simeq -d\sigma_{g_2} -d \sigma_{g_0}\, ,
\eeq
with
\beq
|S^{(g)}|^2 = \mathscr{S}_{\mu\nu} \Pi^{(g)}_{\mu\nu\mu'\nu'} \mathscr{S}^*_{\mu'\nu'}
=-\sum_j c_j |S^{(j)}|^2
,\qquad
\Pi^{(g)}_{\mu\nu\mu'\nu'} =-\sum_j c_j \Pi^{(j)}_{\mu\nu\mu'\nu'}
\eeq
as obtained from the 
squared scattering amplitude
$\mathscr{S}_{\mu\nu}$ summed over
the polarization sum determined by the
4-derivative propagator in the limit $s\ll M_{0,2}^2$.


\subsection{The gravi-ghost width from $2\to 2$ scattering}\label{2->2}
A $2\to 2$ cross section $AB\to R \to A'B'$ between (for simplicity) massless particles and
mediated at tree level in $s$-wave
by a generic resonance $R$ with spin $j$ (multiplicity $g_R =2j+1$) and mass $M$ can be written as
\beq \label{eq:2->2}
 \sigma(A B\to R \to A'B') = {16\pi}\frac{g_R}{g_Ag_B} 
\frac{\Gamma_{R\to AB}\Gamma_{R\to A'B'}}{|k^2 - M^2 + i k \Gamma|^2}.\eeq
Specializing the generic gravi-ghost formula to the $2\to 2$ case determines the decay widths,
in view of their symmetric appearance in eq.\eq{2->2}.


Explicit evaluation then gives, in a theory with $N_s$ real scalars, $N_f$ Weyl fermions, $N_V$ vectors
(all massless or much lighter than $M_{0,2}$):
\beq  \Gamma_2 =- \pi  M_2\frac{f_2^2 }{(4\pi)^2}  \bigg(\frac{N_s}{120}+\frac{N_f}{40}+\frac{N_V}{10}\bigg),\qquad
\Gamma_0  = \frac{\pi M_0 f_0^2}{24(4\pi)^2}\sum_S (1+6 \xi_S)^2.
\label{eq:Gamma02}\eeq
Direct computation shows that, at tree level,
$\Gamma(g_2 \to g g)=\Gamma(g_2 \to g g_0)=0$
(see also~\cite{1804.04980,1806.03605}), and that  $\Gamma(g_2\to g_0 g_0)$
equals 
the decay width into a real scalar with mass $M_0$ (see eq.\eq{g2S}).
In the Standard Model $N_s=4$, $N_f=45$, $N_V=12$ so
$ \Gamma_2 \approx -16\eV \left(\sfrac{M_2}{10^{10}\GeV}\right)^3$.
In cosmology ghosts are not in thermal equilibrium at $T \circa{>} M_2$.

\subsection{The gravi-ghost width from the imaginary part of its propagator}
We start from the simpler case of a non-tachionic 4-derivative scalar containing a normal scalar with mass $m_1$
and a ghost with mass $m_2>m_1$.
The tree-level kinetic term (quadratic part of the action) must be 
\beq \Pi(k) = -(k^2-m_1^2)(k^2-m_2^2), 
\eeq
where $k$ is the quadri-momentum.
Indeed,  assuming for simplicity $m_2\gg m_1$, at low $k^2 \ll m_2^2$ this reduces to  $  \Pi \simeq  m_2^2 (k^2 - m_1^2)$,
which is the usual kinetic term, up to the overall positive normalization factor $m_2^2$.
This shows that the $k^4$ term must have negative sign.

Loop corrections due to interactions with matter generate a positive imaginary part,
${\rm Im}\, \Pi \ge 0$, in view of the  optical theorem.
Ignoring the $k^4$ term in the limit $k^2 \ll m_2^2$,
this means that a normal scalar has a positive width,
$\Pi \simeq m_2^2(  k^2 - m_1^2 + i  m_1 \Gamma_1)$, as well known.
In the opposite $k^2\gg m_1^2$ limit, this means that the ghost has a negative width,
$\Pi \simeq -k^2(k^2 - m_2^2 + i m_2 \Gamma_2)$ with $\Gamma_2\le 0$.
This sign, following from general considerations, agrees with
explicit resummation of the quantum corrections to a ghost propagator~\cite{0902.1585}.

\medskip

Coming to the gravi-ghost case, 
one-loop corrections due to normal matter contribute in a way which  respects
the symmetries of the Lagrangian:
each term gets a correction factor $Z$
\beq \label{eq:Leff}
 \sqrt{\det g}  \bigg[
Z_0 \frac{R^2}{6f_0^2}+ Z_2 \frac{\frac13 {R^2} - R_{\mu\nu}^2}{f_2^2} - \frac{\bp^2}{2} Z_M R
\bigg] . \eeq
Massless matter in the loop give $Z_M=1$ and
\beq
Z_2 =1 - f_2^2  \bigg(\frac{N_s}{120}+\frac{N_f}{40}+\frac{N_V}{10}\bigg)  B_0(k,0,0),\qquad
Z_0  = 1+ \frac{f_0^2}{24}\sum_S (1+6 \xi_S)^2 B_0(k,0,0)
\eeq 
where 
\beq
B_0(k,0,0) = \frac{1}{(4\pi)^2} \bigg[\frac{1}{\epsilon} +\ln(-\frac{\bar\mu^2}{k^2}) + 2\bigg]
\eeq
is the usual Passarino-Veltman function in $d=4-2\epsilon$ dimensions.
Its $1/\epsilon$ pole reproduces the RGE for $f_{0,2}$~\cite{agravity}.
Inverting the gravi-ghost kinetic term coming from the action of eq.\eq{Leff}
\beq \label{eq:gKin1}
\bigg(\frac{k^4}{c_2 f_2^2} Z_2 + \frac{\bar M_{\rm Pl}^2 k^2}{4} Z_M \bigg) \Pi^{(2)}_{\mu\nu\rho\sigma}  +\bigg( \frac{k^4}{c_0 f_0^2} Z_0 - \frac{\bar M_{\rm Pl}^2 k^2}{2} Z_M\bigg)\Pi^{(0)}_{\mu\nu\rho\sigma} \eeq
gives the propagator of eq.\eq{Pgrav} with
\beq \label{eq:propZ}  P^{(j)}(k^2) =  \frac{1}{Z_j k^4 - M_j^2k^2 Z_M}
\qquad \hbox{i.e.}\qquad
\Gamma_j = k\, \Im Z_j
\eeq
which on-shell reproduces the widths in eq.\eq{Gamma02}
inserting 
the imaginary part  $\ln(-1) = i \pi$ in agreement with the optical theorem.


\section{Specific processes}\label{specific}
We now make the previous section more concrete computing specific processes.

\subsection{Decays}
The gravi-ghost decay squared amplitude into a complex scalar with mass $m_S$ is
\beq\label{eq:D0D2}
|D^{(2)}|^2 =
\frac{(s_g-4m_S^2)^2}{120},\qquad
|D^{(0)}|^2 = \frac{[s_g(1+6\xi_S)^2+2 m_S^2]^2}{12}\eeq
such that the decay widths at rest are
\beq  \label{eq:g2S}
\Gamma(g_2 \to SS^*) =- \frac{\pi M_2 f_2^2}{60(4\pi)^2}\Phi^{5/2},\qquad
\Gamma( g_0 \to SS^*)=\frac{\pi M_0 f_0^2(1+6\xi_S + 2 m_S^2/M_0^2)^2}{12(4\pi)^2}\Phi^{1/2}
,\eeq
where $\Phi=1-4\sfrac{m_S^2}{M_{2,0}^2}$.
The gravi-ghost decay rates into a Dirac fermion with mass $m_f$ is
\beq \Gamma(g_2 \to f\bar f) =- \frac{\pi M_2 f_2^2}{20(4\pi)^2}
\bigg(1 +\frac{8m_f^2}{3M_2^2}\bigg)  \Phi^{3/2}
,\qquad
\Gamma( g_0 \to f\bar f)=
\frac{\pi m_f^2 f_0^2 }{6(4\pi)^2M_0} \Phi^{3/2}.\eeq
where now $\Phi=1-4\sfrac{m_f^2}{M_{2,0}^2}$.
The spin-0 component decays at tree level only if conformal invariance is broken by
$\xi_S\neq - 1/6$ or by particle masses $m_S$ or $m_f$~\cite{Salvio:2017qkx}.

\subsection{$2\to 2$ gravitational scattering among  scalars}
To start we consider the gravitational scattering $SS^* \to S'S^{\prime *}$ among two different complex massless
scalars, such that only $s$-wave graviton exchange contributes:
\beq \label{eq:sigmaScal}
\frac{d\sigma}{dt} = \frac{1}{2304\pi  s^2}\bigg|{f_2^2}  (s^2+6ts+{6t^2})P^{(2)} (s)  -
 {f_0^2} \,s^2 (1+6\xi_S)(1+6\xi_{S'})P^{(0)}(s)\bigg|^2,\eeq
 where 
 $t\equiv  (p_1-q'_1)^2$.
The interference between the spin 0 and 2 components cancels in the total cross section,
reproducing the general formul\ae{} of section~\ref{1g}
\begin{eqnsystem}{sys:sigmaSS}
\sigma &=& \frac{s^3}{11520\pi}\bigg[f_2^4 |P^{(2)}|^2 +{5f_0^4}|P^{(0)}|^2 (1+6\xi_S)^2(1+6\xi_{S'})^2\bigg]\\ &\simeq&\left\{\begin{array}{ll}
 s \big[ 1+  5 (1+6\xi_S)^2(1+6\xi_{S'})^2\big]/{2880\pi \bar M_{\rm Pl}^4},
& s\ll M^2_{0,2}, \\[2mm]
 80\pi^2\Gamma_2 \hbox{BR}_i \hbox{BR}_f \delta(s-M_2^2) /M_2,& s\simeq M_2^2,  \\[2mm]
 80\pi \hbox{BR}_i \hbox{BR}_f /M_2^2, & s= M_2^2, \\[2mm]
 \big[ f_2^4+  5f_0^4 (1+6\xi_S)^2(1+6\xi_{S'})^2\big]/{11520\pi s}, 
& s\gg M_{0,2}^2.
\end{array}\right.
\end{eqnsystem}
For $\sqrt{s}\ll M_{0,2}$ the cross section reduces to the Einstein limit,
while for $\sqrt{s}\gg M_{0,2}$ exhibits the softer behaviour typical of renormalizable
dimensionless theories.
Even if $g_0$ and/or $g_2$ are light enough to be kinematically accessible at present colliders, the
cross sections are Planck-suppressed and thereby negligible except close to the pole.
At the peak $\sigma$ saturates  the unitarity bound,
where $\hbox{BR}_{i(f)} $ is the branching ratio into the initial (final) state.
However, a maximal resonant enhancement needs beams with
energy resolution $\Delta$ comparable to the decay widths $|\Gamma_{0,2}|$, otherwise
$\sigma \sim M_2^2/M_{\rm Pl}^2  \Delta^2$.

\subsection{$2\to 2$ scattering among fermions}
The gravitational cross section for $f\bar f \to f'\bar f'$ where
$f$ and $f'$ are  massless different Dirac fermions is
   \beq\frac{d\sigma}{dt} =f_2^4
\frac{ \left(s^4+10 s^3 t+42s^2 t^2+64 s t^3+32 t^4\right)}{2048 \pi 
   s^2}|P^{(2)}(s)|^2.\eeq
The total cross-section reads
   \beq \sigma= \frac{f_2^4 s^3}{5120 \pi }|P^{(2)}(s)|^2.\eeq
Colliders use charged fermions, such that 
the gauge/gravity interference is more important than the above purely gravitational term.

\subsection{$2\to 3$ gravitational scattering among  scalars}\label{SS->Vg}
In the above $2\to2$ scatterings the quadri-momentum $k$ of the gravi-ghost was fixed by kinematics,
$p_1 + p_2 = k = q'_1+q'_2$.
We now consider processes where $k$ is free, allowing to better probe the gravi-ghost behaviour.
We compute the scattering 
\beq \label{eq:SS->Vg}
S(p_1) + S^* (p_2) \to \gamma(q_1) +  [g(k) \to S'(q'_1) + S^{\prime *}(q'_2)]
\eeq
which involves
one massless vector $\gamma$ coupled to a  massless initial-state scalar $S$ with charge $e$,
and a gravi-ghost that transforms into neutral massless scalars $S'S^{\prime *}$.
The  $SS^* \to \gamma g$ scattering amplitudes are
\beq |S^{(2)}|^2 =e^2\bigg( \frac{tu}{s}+s_g+ \frac{s_g^2 s}{6tu}\bigg),\qquad
|S^{(0)}|^2 =e^2 \frac{s s_g^2}{3tu}(1+6\xi_S)^2.\eeq
The decay amplitudes are given in eq.\eq{D0D2}.
In addition to the usual logarithmic IR enhancement,
there is a new kind of IR enhancement in the spin-2 sector 
that arises when the
gravi-ghost has small $k^2 = s_g  \ll s$ (while the individual components of its quadri-momentum
$k_\mu$ can be large), which arises
because the intermediate gravi-ghost
squared propagator has 4 powers of momentum.
Assuming, for simplicity, a scattering squared energy $s$ so high that 
terms suppressed by powers of $s_g/s$ can be neglected,
the cross section mediated by the spin-2 gravi-ghost is
\beq \label{eq:2->3S}
\sigma(SS^*\to \gamma S'S^{\prime*}) \simeq  \int_0^s \frac{ds_g}{\pi} 
\bigg[\frac{f_2^2 e^2}{8\pi}\bigg]
  |P^{(2)}(s_g)|^2  \bigg[\frac{f_2^2}{8\pi}  \frac{s_g^2}{120}\bigg]
\simeq
\frac{e^2 f_2^4}{46080\pi^3} \int_0^s \frac{ds_g}{|s_g - M_2^2 + i M_2 \Gamma_2|^2}.  
 \eeq
The new IR divergence is qualitatively similar to the usual  soft/collinear divergences\footnote{Which are also present since we assumed massless scalars,
and affect the terms subleading in $s_g\ll s$.}
with quantitative differences:
\begin{itemize}
\item[i)] a $s_g^2$ suppression from the decay interaction, which contains two powers of momentum, 
as typical of spin-2 interactions;
\item[ii)] a higher enhancement from the 4-derivative propagator.
\end{itemize}
As a result the new IR enhancement is power-like rather than logarithmic.
In the massless limit of agravity this gives a IR divergence, as expected given that
in this limit the gravitational potential grows with the distance.
In the massive theory, the ghost mass $M_2$
cuts the divergence such that the gravi-ghost-mediated $2\to 3$ cross section is
saturated by the ghost peak $s_g \approx M_2^2$.
Thereby the IR enhancement can be described by  the cross section for
on-shell gravi-ghost production in the $S S^* \to \gamma g_2$ sub-process:
\beq
\sigma(S S^* \to \gamma g_2) = -\frac{f_2^2}{s M_2^2} 
 \int d\Phi_{\rm scattering} |S^{(2)}|^2 \stackrel{M_2^2\ll s}\simeq -\frac{e^2f_2^2}{48\pi M_2^2} =-
 \frac{e^2}{24\pi \bar M_{\rm Pl}^2}.
\eeq
This negative cross section for ghost production, combined with the negative ghost
decay width, yields positive cross sections among matter particles.
The cross section for producing the massless graviton equals
$\sigma(S S^* \to \gamma g)= \sfrac{e^2}{24\pi \bar M_{\rm Pl}^2}$. 
Above the peak at $s_g = M_2^2$ the two processes interfere negatively,
reducing the differential cross section $d\sigma/ds_g$.

\subsection{$2\to 3$ gravitational scattering among fermions}\label{ee->Vg}
We consider $e\bar e \to \gamma [g\to\nu\bar\nu]$, which is
the fermionic analogue of the scalar process computed in section~\ref{SS->Vg}:
scattering of two charged fermions (electron-positron) into a vector (photon) and a gravi-ghost,
that decays into two neutral fermions (neutrinos).
The $2\to2$ purely gravitational
scattering sub-amplitudes are, in the limit $m_e=0$:
\beq |S^{(2)}|^2 =e^2\left[
\frac{s_g (s_g^2+s^2)}{16 t u}+\frac{2 (s_g^2+ s^2)-s s_g-4 tu}{8 s}\right]
,\qquad
|S^{(0)}|^2 =0.\eeq
The cross section mediated by the spin-2 component in the limit $s_g\ll s$ is
\beq \sigma(e\bar e\to \gamma \nu\bar\nu) = \int_0^s \frac{ds_g}{\pi} 
\bigg[\frac{f_2^2 e^2}{48\pi}\bigg]
 \times  |P^{(2)}(s_g)|^2 \times \bigg[\frac{f_2^2 s_g^2}{320\pi}\bigg]
 \eeq
which has a structure analogous to the scalar case in eq.~\eq{2->3S}. 
%
The cross section for on-shell  ghost production is
\beq \sigma(e\bar e\to \gamma g_2) \stackrel{s\gg M_2^2}\simeq-
\frac{e^2 f_2^2}{48\pi M_2^2}\eeq
again equal (up to a sign) to the
cross section for graviton production.


\subsection{Production of two gravi-ghosts: $SS^*\to g_{j_1}g_{j_2}$}
In the limit $s,|t|\gg M_2^2$ and $\xi_S=-1/6$ we find 
\beq |S^{(2,2)}|^2 \simeq \frac{t^2 (s+t)^2}{8s^2}
 .\eeq
The total $S S^*\to g_2 g_2$ cross section is doubly IR enhanced, and grows with $s$: 
\beq 
\sigma_{g_2g_2} \stackrel{s\gg s_g}{\simeq} \frac{f_2^4 s}{960\pi M_2^4}
= \frac{s}{240\pi \bar M_{\rm Pl}^4}.
\eeq
In the same limit $\sigma_{g_2g_2}\simeq \sigma_{gg} \simeq -\sigma_{gg_2}\simeq -\sigma_{g_2g}$.
The cross section for the production for the scalar components of the graviton $g_0$ behaves
as the cross section for production of a scalar~\cite{Salvio:2017qkx}:
$\sigma_{g_0 g_2}=0$ and
\beq |S^{(0,0)}|^2 \simeq M_2^4\frac{(s^2+6st+6t^2)^2}{144s^4},
\qquad
\sigma_{g_0g_0}\simeq \frac{f_2^4}{11520\pi s}
.\eeq

\begin{figure}
\begin{center}
\includegraphics[width=0.95\textwidth]{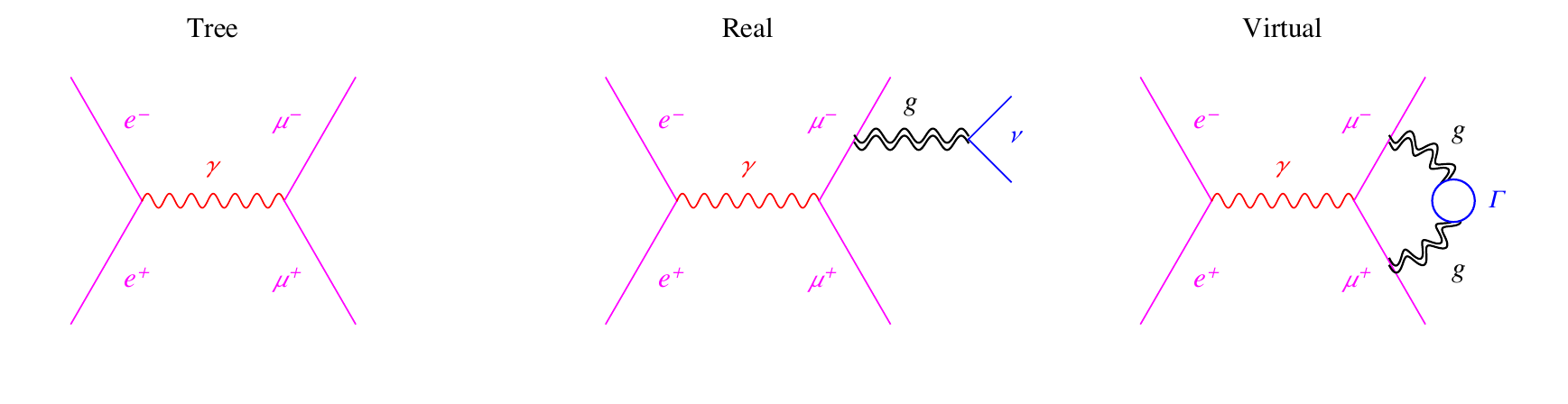}
\caption{\em \label{fig:realvir} A sample process.
The second diagram, computed with real (Minkowskian) moments, has an IR power enhancement
at large gravi-ghost momenta, $k =(E,E,0,0)$.
\label{fig:agv}}
\end{center}
\end{figure}

\section{Ghostrahlung}\label{Ghostrahlung}
IR and collinear enhancements and divergences are a well studied topic in QED, QCD and gravity.
Based on previous experience, 
we  explore the  implications of the new IR enhancements present in 4-derivative theories.
In section~\ref{soft} we discuss the consequences of soft theorems.
In section~\ref{analytic} we try to go beyond the soft limit.
In section~\ref{Faddev} we consider the approach by Kulish and Faddeev.
Based on previous discussions, 
in section~\ref{physics} we draw our conclusions.

To make the discussion more concrete we compute a toy process $S \to f\bar f$:
a heavy scalar $S$ at rest with mass $m_S \gg M_2$ decays into massless fermions. 
Adding one  gravi-ghost with energy $E$ to the final state, 
its energy distribution in terms of $x=2E/m_S$ is
\beq \label{eq:Sffg}
\frac{dN_{g}}{dx} =\frac{2f_2^2 m_S^2}{3(4\pi)^2M_2^2} \frac{1-7x/4+9x^2/8-3x^3/8}{x },\qquad
\frac{dN_{g_2}}{dx} =-\frac{dN_{g}}{dx}\sqrt{1-\frac{4M_2^2}{xm_S^2}}\eeq
for the graviton and ghost components, respectively.
We only wrote the contribution enhanced by $m_S^2/M_2^2$, which is the new IR  enhancement.
The usual IR enhancement produced by the fermion propagator gives the $1/x $ factor.

\subsection{Soft theorems}\label{soft}
At the diagrammatic level, {soft theorems} capture IR enhancements in terms of the
behaviour of couplings and propagators.
In the usual 2-derivative case,
emission of spin 1 photons and gluons
is enhanced by soft  and collinear logarithmic divergences,
while emission of spin-2 gravitons is only enhanced by soft logarithmic divergences, because
graviton couplings are suppressed
as $\theta^2$ in the small-angle limit $\theta\ll1$,
canceling collinear divergences in the propagator~\cite{Weinberg:1965nx,1109.0270,1207.4926}.
The main result of such techniques is that IR divergences cancel 
between `real' and `virtual' corrections
when  computing appropriate `IR-safe' observables, which discount too soft particles as unobservable.
Indeed, enhanced soft/collinear radiation arises from virtual particles which are
almost on shell  and propagate for long time, invalidating the difference between real and virtual particles.

\subsubsection*{IR enhancements in 4-derivative gravity}\label{unknown}

IR enhancements are stronger in 4-derivative theories  because
propagators of massless or light fields with small momentum $k_\mu$ are
more IR-divergent: $1/k^4$ rather than $1/k^2$.
This comes together with the extra issue of understanding what a `ghost' is,
that we addressed by treating the gravi-ghosts as virtual particles in scatterings among matter particles,
finding that 4-derivative gravity contains {power-like} IR enhancements
at small $k^2$.
This includes  the  region of small $k_\mu$ where soft theorems hold, but goes beyond it, 
including the region where the gravi-ghost splits into collinear ultra-relativistic  matter particles.

Adding the Einstein term, the ghost acquires a mass $M_2$, and IR divergences
get replaced by IR enhancements, saturated around the ghost pole, $k^2 \simeq M_2^2$. 
This is the region where it is more difficult to understand what a ghost is,
given that understanding any massive particle needs a non-perturbative resummation of
higher-order corrections to their propagator, that acquires Breit-Wigner form.

\subsubsection*{Soft region}
It is partially useful to explore the implications of
soft theorems, which control a part of the new IR enhancement.
The amplitude $\mathscr{A}_{\mu\nu}$ 
for emitting a soft graviton with small momentum $k_\mu$
in a  process among matter particles
with amplitude $\mathscr{A}_{\rm hard}$ and momenta $p_i \gg k$
has factorised form~\cite{Weinberg:1965nx}
\beq \label{eq:softJ}
\mathscr{A}_{\mu\nu} = \mathscr{A}_{\rm hard} J_{\mu\nu},\qquad J_{\mu\nu}(k)=
  -\sum_i \frac{p_{i\mu} p_{i\nu}}{2p_i\cdot k}.\eeq
All momenta are here in-going, so $p_i = -q_i$ for
final-state particles with out-going momentum $q_i$.
Summing the squared amplitude  over gravi-ghost polarizations gives
\beq 
J_{\mu\nu}(k)\Pi^{(j)}_{\mu\nu\mu'\nu'}(k) J^*_{\mu'\nu'}(k)\stackrel{k\to 0}\simeq |J^{(j)}|^2 \, ,
\eeq
where
\beq |J^{(2)}|^2\equiv  \sum_{ij} 
\frac{(p_i\cdot p_j)^2-p_i^2 p_j^2/3}{4(p_i\cdot k)(p_j\cdot k)},\qquad
|J^{(0)}|^2\equiv\sum_{ij} 
\frac{p_i^2 p_j^2/3}{4(p_i\cdot k)(p_j\cdot k)},\qquad
|J^{(g)}|^2  =-\sum_{j=\{0,2\}} c_j J^{(j)}
\eeq
up to terms finite for $k\to 0$.
The cross section among matter matter particles mediated
by a virtual gravi-ghost is given in the soft limit by eq.\eq{sigma23} with 
\beq
|S^{(j)}|^2 \stackrel{k\to0}\simeq |\mathscr{A}_{\rm hard} J^{(j)}|^2.\eeq 
Eq.\eq{sigmagj} then gives the cross section $\sigma_{g_j}$ for producing one 
on-shell soft gravi-ghost $g_j$ in terms of
the cross section $\sigma_{\rm hard}$ without gravi-ghosts:
%
\beq \label{eq:realj}
d\sigma_{g_j}  \stackrel{k\to0}\simeq  d\sigma_{\rm hard}  \frac{d^3k}{(2\pi)^3 2E_j} \frac{2c_j }{\bp^2}
|J^{(j)} |^2 \, ,
\eeq
where the $J^{(j)}$ factors contain the usual IR enhancement.
Applied to our toy process,
soft theorems ($|J^{(2)} |^2= 2/3x^2$ adding a soft graviton to a $1\to 2$ decay with massless final-state particles) correctly reproduce the leading $1/x$ term of 
eq.\eq{Sffg} for $x\ll1$.\footnote{Even the sub-leading term is captured from the soft formula, if one keeps the $k^2$ in the eikonal propagators.
For graviton emission from a scattering involving scalars only, this reproduces the full amplitude, because
non-soft terms with $k_\mu$ at the numerator in the amplitude 
must be contracted with graviton polarizations, and thereby vanish.
Then IR enhancements cancel in the total cross section.
Processes involving fermions or vectors contain different contractions with their polarizations,
so that the soft limit fails at $x\sim 1$.}

\subsubsection*{Cancellation of real with virtual corrections in the soft limit}
We now show that, in the soft limit, the real IR enhancement gets canceled by virtual corrections.
The total amplitude for emission of one soft graviton is given by eq.\eq{softJ}.
Its IR divergence gets  canceled by the virtual correction $\delta  \mathscr{A}_{\rm hard}$,
obtained by considering the amplitude for emission of two soft
gravi-ghosts, 
$\mathscr{A}_{\mu\nu\alpha\beta}=\mathscr{A}_{\rm hard} J_{\mu\nu}(k)J_{\alpha\beta}(-k)$
and connecting them into a one-loop diagram through a gravi-ghost propagator.
(The cubic graviton vertex does not contribute to the leading IR divergence).
In 4-derivative gravity one gets
\beq \label{eq:virtual}
\frac{\delta  \mathscr{A}|_{\rm hard}}{\mathscr{A}_{\rm hard}} \stackrel{k\to0}\simeq \frac12 \int \frac{d^4k}{(2\pi)^4} P_{\mu\nu\alpha\beta} J_{\mu\nu}(k)J_{\alpha\beta}(-k)
=-\frac12
\sum_{j=0,2} \int \frac{d^4k}{(2\pi)^4}  i c_j f_j^2 P^{(j)}(k) |J^{(j)}|^2.
\eeq
In order to match with the real correction of eq.\eq{realj} 
we perform the integral over $dk_0$.
Keeping generic signs of the $i\epsilon$ in the two components of the
4-derivative propagator gives
\beq \label{eq:dk0}
\int_{-\infty}^{+\infty} \frac{dk_0}{2\pi}  \frac{i}{[k_0^2-E^2\pm  i \epsilon][
k_0^2-E^{\prime 2} \pm' i \epsilon]} \stackrel{\epsilon\to 0}\simeq \frac{1}{E^{\prime 2}-E^2}\bigg(\frac{\pm'}{2E^\prime}- \frac{\pm}{2E}\bigg).\eeq
The  virtual correction decomposes as the sum
over the massless graviton ($E_g=k$), the ghost $(E_2^2 = k^2 + M_2^2$) and the
spin 0 component ($E_0^2 = k^2 +M_0^2$):
\beq \frac{\delta  \mathscr{A}|_{\rm IR}}{\mathscr{A}} \stackrel{k\to0}\simeq -\frac12
\int \frac{d^3k}{(2\pi)^3}\frac{2}{\bp^2} \bigg[\pm
\frac{|J^{(g)}|^2}{2E_g}
 \pm' c_2 \frac{|J^{(2)}|^2}{2E_2}
 \pm' c_0\frac{|J^{(0)}|^2}{2 E_0}\bigg] .
\eeq
For the $i\epsilon$ prescription that makes the theory renormalizable,
$\pm=\pm'=+$, the
virtual soft correction cancels the real soft  correction of eq.\eq{realj}.\footnote{`Real' and `virtual' effects can be unified viewing {squared amplitudes as 
imaginary parts} of higher loop diagrams.
Each higher diagram is separately IR convergent, so that the cancellations mentioned above
(at the level of the total amplitude) take  part separately between the cuttings of each higher diagram.}

The above leading-order cancellation, diagrammatically illustrated in fig.~\ref{fig:realvir},
persists at higher orders, where diagrams with a 
series of matter bubbles on the gravi-ghost propagator give the dominant IR effect.
However, resumming corrections to the gravi-ghost propagator transforms it into a
Breit-Wigner with a negative width $\Gamma_2 <0$,
which shifts the ghost pole to the acausal region, behaving as a wrong-sign $i \epsilon$ prescription.
As a consequence
the loop integral of the resummed propagator is not the resummation of the loop integrals:
one corresponds to $\epsilon > |\Gamma| \to 0$, the other to $|\Gamma| >\epsilon\to 0$.


\subsection{Beyond the soft limit}\label{analytic} 
The $J$ factors in eq.\eq{virtual}
give the usual IR 
enhancements, and the 4-derivative propagator gives the
new IR enhancement.
Focusing on it, one can ignore the $J$ factors and the external momenta.
Then, virtual corrections have the typical form of dimensionless theories,
exemplified by dimensionless loop integrals such as $\int d^4k/k^4$.
In the Euclidean they lead to logarithmic divergences, 
that induce RGE running of the couplings.
Power-like IR enhancements appear in the  Minkowskian,
as clear from
$d^4k/k^4 = ds_g/s_g^2 \times d^3k/2E$: the integral over
$s_g = k^2$ is power divergent at $s_g\to 0$
(and enhanced in the presence of masses).

Non-soft IR enhancements arise from the extra region
where $s_g=k^2$ is  small, while the individual components of $k_\mu$ are large.
This is a collinear configuration (although different from collinear enhancements in
2-derivative theories), which only exists in the Minkowskian.
Collinear divergences do not exist in the Euclidean,
where small $k_E^2>0$ implies that each component of $k_E$ is individually small.

IR enhancements beyond the soft limit would get under better
control if one could use Euclidean techniques.
However ghosts complicate the connection between the Minkowskian and Euclidean theory.
One may need an integration contour deformed to lie above the ghost pole, 
or something along the lines of~\cite{Coleman} or~\cite{1703.04584}.

\subsection{Asymptotic scattering states}\label{Faddev}


The physics of IR divergences has been clarified by Kulish and Faddeev~\cite{Faddeev},
with recent developments~\cite{1308.6285,1608.05630,1705.04311,1803.02370}.
IR divergences appear when the LSZ formula for cross sections
(which assumes that particles are free at asymptotically large distances) 
is used despite being unapplicable due to the presence of long-range interactions.

2-derivative theories give Coloumbian-like interactions, which do not induce
IR divergences provided that cross sections are computed among 
scattering states that account for such long-range interactions.
The correct scattering states are well known in non-relativistic quantum mechanics.
The appropriate relativistic states have been computed in QED~\cite{Faddeev},
and look like a charged particle surrounded by a cloud of soft photons.
This approach allows to define a $S$-matrix, not just some IR-safe observables.


\medskip

The Coulomb  force $e^2/r^2$ vanishes at large distances, giving a soft IR divergence. 
Agravity in the massless limit produces a constant gravitational force $\sim f_2^2 s $, corresponding to a non-soft IR divergence.
Since the force is confining, the only possible scattering states have zero energy~\cite{Boulware:1983td}.
When the Einstein term is included, agravity at large distances $r \circa{>}1/M_{0,2}$ predicts to the usual Newton force, which behaves as a Coulomb force with coupling $E/M_{\rm Pl}$.
The associated soft IR divergence changes character when the coupling becomes large, 
namely in super-Planckian scatterings.

\subsection{Discussion}\label{physics}

Based on the previous discussion, we draw the physical conclusions.

Tree-level $2\to 2$ gravity-mediated cross sections 
are of order $\sigma \sim f_{0,2}^4/s$ in the agravity regime  $\sqrt{s}\gg M_{0,2}$.
They are suppressed by the agravity
couplings $f_{0,2}$, that we assume to be small in order to keep the Higgs mass
naturally smaller than the Planck mass~\cite{agravity}.

However, cross sections get infra-red enhanced by powers of $s/s_g$
when more gravi-ghosts are involved, such that a virtual gravi-ghost with momentum $k_\mu$
can have $s_g \equiv k^2 \ll s$.
The largest cross section is obtained adding gravi-ghost interactions to a
$2\to 2$ scattering mediated by order-one matter couplings,
like those present in the Standard Model.
In such scatterings, the leading-order
cross sections for emitting a graviton or a ghost are as large as in Einstein gravity,
and violate unitarity bounds at $\sqrt{s}\circa{>}M_{\rm Pl}$.

The total number of radiated gravitons is well approximated by soft limits.
Resummation of soft radiation shows that it leaves
total cross sections roughly unchanged
in view of cancellations between real and virtual effects~\cite{Weinberg:1965nx}.

\smallskip

The total radiated energy is instead dominated by hard gravitons, with $E \sim\sqrt{s}$.

\medskip

%

In Einstein gravity,
large cross section arise  because of UV divergences: the theory is non-renormalizable
and Planckian gravitons are strongly coupled.
Thereby, after being emitted, they re-scatter giving rise to complicated 
higher-order phenomena,
in particular formation of black holes 
which has been conjectured to lead to classicalization
 (see~\cite{1409.7405,1512.00281} for recent studies).

\smallskip

Agravity is renormalizable and the same tree-level cross section arise because of IR enhancements.
The key physical difference with Einstein theory
is that (super)Planckian gravitons
are weakly coupled: after being radiated they simply carry away their energy.
This is why, to focus on the largest most problematic cross section,
we gravitationally dressed scatterings 
mediated by order-one matter couplings, rather than by gravity.
The fraction of the total energy that is radiated to gravi-ghosts is $\sim B/(1+B)$, where
$B \sim f_2^2 {s} /(4\pi M_2)^2 \sim s/(4\pi M_{\rm Pl})^2$
is the soft emission factor of~\cite{Weinberg:1965nx}.
This implies that, because most of the energy is lost to gravi-ghost radiation, 
super-Planckian scatterings
get effectively down-graded to Planckian, which have cross sections within unitarity bounds.
The new IR enhancement is non-soft: the non-cancellation of real and virtual corrections allows for
large corrections to the cross sections.

The super-Planckian energy is radiated away by hard gravi-ghosts which are almost free.
Agravity provides a perturbative classicalization mechanism, while non-perturbative black holes play a negligible role around the Planck scale.
Indeed the energy content of gravitational fields is smaller in agravity than in Einstein gravity.
This can be estimated as the Newton potential evaluated at $r \sim 1/M_{0,2}$,
and can be precisely computed in specific cases: for example 
the one-loop gravitational correction to the mass $M$ of 
a Dirac fermion is $ \Delta M = \sfrac{5 f_2^2 M^2}{96\pi M_2}$,
indicating that Einstein black holes arise in agravity at  $M_{\rm BH} \circa{>} M_{\rm Pl}/f_2$~\cite{agravity,1502.01028}.

\section{Conclusions}\label{end}
4-derivative gravity is renormalizable thanks to extra
graviton components, especially a spin-2 ghost with mass $M_2$,
to be quantised with positive energy.
We computed the ghost behaviour {\em \`a la} Lee-Wick: by viewing ghosts
as virtual particles that mediate scattering among ordinary matter particles.

We restricted our attention to tree-level processes,
which teach a significant amount of physics, 
and which 
are not affected by details of quantisations proposed to make sense of ghosts at higher orders.

Some cross section have the expected good behaviour typical of renormalizable interactions,
being softer than analogous cross sections in Einstein theory.
Cross sections for on-shell ghost production
can be enhanced colliding beams with energy spread as small as the ghost decay width
$\Gamma_2 \sim -M_2^3/M_{\rm Pl}^2$.
The negative sign signals  micro-acausality, which can be probed by
an observer at large distance through a $2\to 2$ process among matter particles 
mediated in the $s$ channel by the ghost, 
measuring that the secondary  vertex is displaced
in the unusual direction, as if the ghost decayed before being produced.

\medskip

However, other cross sections where gravi-ghosts have small virtuality
do not have the expected behaviour typical of renormalizable interactions.
Interpreted in terms of real gravi-ghosts, we find that
tree-level cross sections for emitting $n$ gravitons grow 
as $(s/M_{\rm Pl}^2)^{n}/s$.
These cross sections are as bad as in Einstein gravity, as they are not affected by the
extra physics present in agravity.
Furthermore, tree-level cross sections for emitting one gravitational ghost behave in the same way.

In Einstein gravity, such non-unitary super-Planckian cross section are
a manifestation of the UV-divergent behaviour
typical of non-renormalizable theories.

In agravity, such large cross section arise because of a new kind of IR enhancement,
typical of 4-derivative theories, and due to the
small-momentum behaviour of a propagator with 4 powers of momentum.
This enhancement is a remnant of IR divergences present in the mass-less limit of agravity,
where emission of each gravi-ghost 
with virtuality $s_g=k^2$ is accompanied by a factor $\int ds_g/s_g^2$.
In Minkowski space, $k^2$ can be small even if the components $k_\mu$ are large,
such that the new IR effect extends beyond the soft region.


\medskip

In Einstein gravity, a super-Planckian scattering is accompanied by large energy losses 
dominated  by hard gravitons with Planckian energy.
Gravitational fields contain most of the energy and Planckian gravitons are strongly coupled:
they re-scatter forming macroscopic black holes, which possibly lead to classicalization.

In agravity, graviton radiation is accompanied by ghost radiation.
More importantly,
the energy in gravitational fields is negligible and the Planckian gravi-ghosts are weakly coupled: 
they merely carry away energy, downgrading scatterings to sub-Planckian
and thereby screening super-Planckian physics.
We argued that a resummation of initial-state gravi-ghost radiation 
can significantly affect the cross-sections (analogously to how QED radiation
induces the radiative return of the $Z$-peak), bringing them down 
within unitarity bounds.
This conclusion is consistent with the Kulish-Faddeev understanding of
IR divergences as an effect of long-range dynamics,
taking into account that
4-derivative gravity gives, in the mass-less limit, a confining gravitational interaction.

\footnotesize

\subsubsection*{Acknowledgments}
The authors thank Rashmish Mishra and Damiano Anselmi for useful discussions. This work was supported by the ERC grant NEO-NAT and by the grants IUT23-6, PUT799, by EU through the ERDF CoE program grant TK133 and by the Estonian Research Council via the Mobilitas Plus grant MOBTT5.

\footnotesize


\footnotesize


\end{document}